\documentclass[preprint,showpacs,preprintnumbers,amsmath,amssymb]{revtex4}

\usepackage{graphicx}
\usepackage{dcolumn}
\usepackage{bm}

\begin{document}


\title{Development of Kolmogorov spectrum in the pulsar radio emission}


\author{G.~Z.~Machabeli}%
\affiliation{Center for Plasma Astrophysics, Abastumani
Astrophysical Observatory, Ave. A. Kazbegi 2, Tbilisi 0160,
Georgia}
\author{G.~Gogoberidze}
\email{gogober@geo.net.ge} \affiliation{Center for Plasma
Astrophysics, Abastumani Astrophysical Observatory, Ave. A.
Kazbegi 2, Tbilisi 0160, Georgia}

\date{\today}

\begin{abstract}

It is shown that the scattering of electromagnetic waves by
Langmuir ones taking into the account the electric drift motion of
particles, is the most intense nonlinear process and should be
responsible for the formation of radiation spectrum of radio
pulsars. Performed analysis indicates that in the case of
existence of inertial interval formation of stationary spectra is
possible. Analysis of linear mechanisms of wave generation and
silk allow to conclude that only possible stationary solution has
spectral index $\alpha =-1.5$. Obtained spectrum is in good
agreement with observational data.

\end{abstract}

\maketitle

\section{Introduction}               

Aim of the present work is to study nonlinear processes
responsible for the formation of the radio emission spectrum in
the pulsar magnetosphere. The spectra of pulsar radio emission has
been intensively studied by observers
\cite{M96,MKKW00,LYLG95,IKMS81, LSG71}. Above 100 MHz observed
intensity $I$ of the majority of pulsars can be described by a
simple power law $I \sim \nu^\alpha$ with average value of
spectral index $\langle \alpha \rangle =-1.8\pm 0.2$
\cite{MKKW00}. Spectral index of individual pulsars vary from
$\alpha=+1.4$ to $\alpha=-3.8$. Some pulsars has the spectra which
can be modelled by the two power law \cite{M96,LYLG95}, but it
seems that single power law is a rule and the two power law
spectrum is a rather rare exception \cite{MKKW00}.

From our point of view pulsar radio emission is generated and
formed in the magnetosphere. Therefore properties of the radiation
is mainly determined by the physical processes in the
magnetosphere plasma. This approach allows to explain other main
characteristics of pulsar radiation, such as polarization
properties \cite{KMM91}, nullings \cite{KMMS96}, micro impulses
\cite{KMMS01}, mode switching \cite{MMMM97} and so on.

In the presented paper we study nonlinear processes in the
magnetosphere plasma that should be responsible for the formation
of the pulsar radio emission spectra. Mathematical methods used
for analysis is based on the weak turbulence theory (see, e.g.,
\cite{ZK78, ZLF}).

The paper is organized as follows: the model of the magnetosphere
as well as linear eigenmodes of $e^-e^+$ plasma and linear
mechanisms of their generation is discussed in Sec. 2. Nonlinear
processes are studied in Sec. 3. In Sec 4 stationary solutions fot
the spectra are obtained and analyzed.

\section{Linear modes and their generation in pulsar magnetosphere}

\begin{figure}[b]
\includegraphics[]{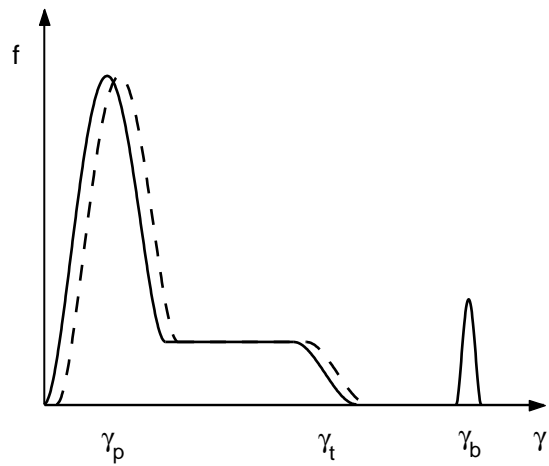}
\caption{\label{fig:fig1} Particle distribution function $f$ of
magnetosphere plasma. Solid and dashed lines correspond to
electrons and positrons respectively.}
\end{figure}

According to the standard model of the pulsar magnetosphere
\cite{GJ69,S71} relativistic $e^-e^+$ plasma that is pierced by
electron beam moves along open field lines of pulsar magnetic
field ${\bf B}_0$. Magnetic field of pulsar is supposed to be
dipole. Under open field lines we mean the field lines that
crosses the surface of light cylinder - the surface where the
velocity of field line solid rotation reaches the speed of light
$c$. It is supposed that electric field directed along the open
field lines is generated at the surface of pulsar \cite{GJ69},
that pulls out electrons of the primary beam from the surface of
the star. The particles moving along magnetic field lines generate
$\gamma$ rays, which on its turn if the energy $\epsilon_\gamma >
2m_e c^2$ generate electron-positron pairs \cite{S71}. This
process is stopped when the initial electric field becomes
screened. The particle distribution function that has the form
presented in Fig. \ref{fig:fig1} can be presented as:
\begin{equation}
f=f_p+f_t+f_b, \label{eq:11}
\end{equation}
where $f_p$ describes the bulk of the plasma, $f_t$ is the
distribution function of elongated to the plasma motion tail
particles and $f_b$ corresponds to the particles of the primary
beam. The distribution function (\ref{eq:11}) is one dimensional.
In Fig. \ref{fig:fig1} solid and dashed lines correspond to
electrons and positrons respectively. The shift of the
distribution functions is caused by the existence of the primary
beam. It is supposed that
\begin{equation}
n_p\gamma_p\approx n_t\gamma_t \approx \frac{1}{2}n_b\gamma_b,
\label{eq:12}
\end{equation}
where $\gamma_p, \gamma_t$, $\gamma_b$ are Lorentz factors and
$n_p, n_t$, $n_b$ concentrations of bulk, tail and beam particles
respectively. For typical pulsars $\gamma_b\sim 10^6-10^7$,
$n_b^0\sim 10^{11}$, $\gamma_t\sim 10^4$, $n_t^0\sim
10^{13}-10^{14}$, $\gamma_p\sim 10$ ¨ $n_p^0\sim 10^{17}$. Index
$0$ indicates that the values belong to the surface of the pulsar.

The condition of quasi neutrality yields
\begin{equation}
\Delta \gamma = \gamma_+ -\gamma_-=\int f_+\gamma d^3p-\int
f_-\gamma d^3p, \label{eq:13}
\end{equation}
here indexes $\pm$ are related to positrons and electrons
respectively. $f_\pm$ are normalized such that $\int f_\pm
d^3p=1$.

In spite of the fact that $\Delta\gamma\ll\gamma_\pm$ this small
difference plays important role in explanation of polarization
properties of pulsar radiation. The radiation of the most pulsars
has considerable part of circularly polarized radiation. It was
shown \cite{KMM91} that circular polarization can occur only in
small angle $\theta$ between wave veqtor ${\bf k}$ and ${\bf
B}_0$:
\begin{equation}
\theta^2 \lesssim \frac{\omega}{\omega_B}\Delta\gamma \ll 1.
\label{eq:15}
\end{equation}
Consequently, the observer receives the radiation that is formed
in small angle with respect to ${\bf B}_0$. So, for pulsar radio
emission the waves propagating nearly along the magnetic field are
important. This agrees with linear theory of $e^-e^+$ plasma in
strong magnetic field: analysis provides that the wave generation
takes place only almost along the magnetic field
\cite{KMM91,VKM85}.

Due to the absence of gyrotropy the spectra of linear modes in
$e^-e^+$ plasma is meager. There exist only three modes
\cite{VKM85}: purely transversal $t$ mode and in general partially
potential $lt$ modes. Electric vector of $t$ mode is perpendicular
to both ${\bf k}$ and ${\bf B}_0$. If ${\bf k}
\parallel {\bf B}_0$ the dispersion relation of this mode is:
\begin{equation}
\omega_t= k c (1-\delta), \label{eq:21}
\end{equation}
where $\delta\equiv\omega_p^2/(4\omega_B^2
\langle\gamma^3\rangle)$, and $\omega_p$, $\omega_B$ are plasma
and cyclotron frequencies respectively
\begin{equation}
\omega_p\equiv \left( \frac{8\pi e^2n_p}{m_e}\right)^{1/2},~~~
\omega_B\equiv \frac{e |B_0|}{m_e c}, \label{eq:22}
\end{equation}
angular brackets denote averaging over distribution function and
$\langle\gamma\rangle\approx\gamma_p$.

$lt$ mode has two branches $lt_1$ and $lt_2$. Electric vectors of
these modes are located in the plane formed by ${\bf k}$ and ${\bf
B}_0$. If ${\bf k} \parallel {\bf B}_0$ mode $lt_2$ merges with
$t$ mode with spectrum (\ref{eq:21}). The mode $lt_1$ propagating
along ${\bf B}_0$ is longitudinal Langmuir wave $(l)$. If $\omega
\gtrsim kc$ their dispersion is \cite{S60,TS61}:
\begin{equation}
\omega_l^2 = \omega_p^2\langle\gamma^{-3}\rangle+ 3k^2c^2\left( 1-
\frac{\langle\gamma^{-5}\rangle}{\langle\gamma^{-3}\rangle}
\right), \label{eq:241}
\end{equation}
and when $\omega \approx kc$:
\begin{equation}
\omega_l \approx c\left[ k- \alpha(k- k_0) \right], \label{eq:242}
\end{equation}
where $k_0^2=2\langle\gamma\rangle\omega_p^2/c^2$ and
$\alpha=\langle\gamma\rangle/2\langle\gamma^3\rangle$.

\begin{figure}[t]
\includegraphics[]{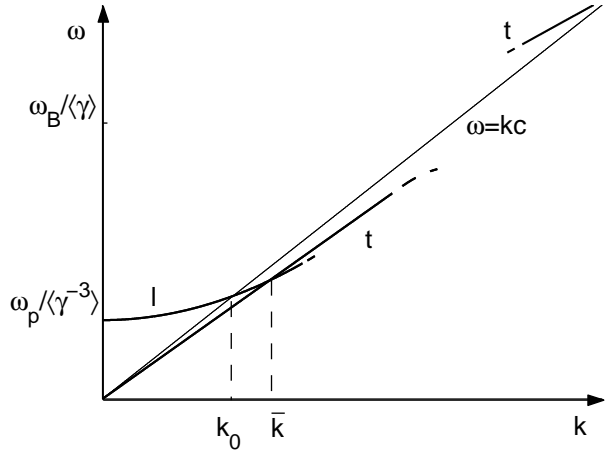}
\caption{\label{fig:fig2} Dispersion curves of the waves
propagating along magnetic field in magnetized $e^-e^+$ plasma.}
\end{figure}

From Eqs. (\ref{eq:21}) and (\ref{eq:242}) it follows that
frequencies of $l$ and $t$ modes become equal at:
\begin{equation}
k \approx \bar k \equiv k_0 \left(1 +
\frac{1}{2}\frac{\omega_p^2}{\omega_B^2}\langle \gamma^3
\rangle\right), \label{eq:243}
\end{equation}
Dispersion relations of the waves propagating along ${\bf B}_0$
are presented in Fig. \ref{fig:fig2}.

Mechanisms of wave generation in the magnetosphere $e^-e^+$ plasma
has been intensively studied (see, e.g., \cite{M86}). The
magnetosphere plasma can be unstable due to two reasons: a)
existence of energetic beam; b) absence of symmetry and one
dimensionality of the distribution function. Detailed analysis
provides \cite{KMM91} that the beam instability can develop only
due to the drift of the particles in the curved magnetic field (so
called "drift-cherenkov" resonance). Linear stage of this
instability was studied in \cite{LMB99} and nonlinear stage in
\cite{SMMK03}. In this paper we will be interested in the second
mechanism of instability - excitation at the anomalous
Doppler-effect resonance. The condition of this resonance is
\cite{KMM91}:
\begin{equation}
\omega-k_\parallel v_\parallel -k_\perp u_\perp^{m}
+\frac{\omega_B}{\gamma_r}\approx 0. \label{eq:25}
\end{equation}
Here $k_\parallel$ and $k_\perp$ are projections of wave vector
along and perpendicular to ${\bf B}_0$, $u_\perp^{m}\equiv c
v_\parallel \gamma_r/\omega_B R_B$ is curvature drift velocity,
$R_B$ is curvature radius of magnetic field lines and $\gamma_r$
is Lorentz factor of resonant particles.

Detailed analysis provides that this condition can be satisfied
only for energetic particles of the tale and beam \cite{LMM79},
when the condition $2\gamma_r^2\delta>1$ is satisfied.
Substituting Eq. (\ref{eq:12}) into (\ref{eq:25}), for $k_\perp=0$
we obtain:
\begin{equation}
\omega_t \approx k_\parallel c \approx
\frac{\omega_B}{\gamma_r\delta}. \label{eq:27}
\end{equation}
From this equation for typical pulsar parameters \cite{MU79,MU89}
it follows that the frequencies of generated $t$ waves
\begin{equation}
\omega_t \ll \omega_B. \label{eq:28}
\end{equation}

It should be noted that transversal waves are strongly damped by
cyclotron damping with the bulk particles. The condition of this
resonance yields
\begin{equation}
\omega-k_\parallel v_\parallel -k_\perp u_\perp^{m}
-\frac{\omega_B}{\gamma_r}\approx 0. \label{eq:26}
\end{equation}
This condition provides that cyclotron damping is effective for
the waves with frequencies
\begin{equation}
\omega_t \gtrsim 2\gamma_p \omega_B. \label{eq:261}
\end{equation}

Linear mechanisms of the wave generation discussed in this section
emerges transversal $t$ waves. On its turn, nonlinear interactions
of  waves redistributes the energy among different modes and
scales and as it will be shown bellow can lead to the formation of
stationary spectra of wave turbulence.

\section{Nonlinear processes}

For the development of weak turbulence theory the small parameter
of weak turbulence is used. In the case of $e^-e^+$ plasma this
parameter is
\begin{equation}
\frac{|E|^2}{m_ec^2n_p\gamma_p}\ll1, \label{eq:31}
\end{equation}
where ${\bf E}$ is electric vector of the wave. This condition
implies that we consider the state when the energy density of
plasma is much greater then the energy density of waves. except
the parameter (\ref{eq:31}) in the magnetosphere plasma there
exist another two small parameters that greatly simplify future
analysis. The first small parameter is
$\omega_p^2/\omega_B^2\ll1$. The second one:
\begin{equation}
\gamma_p^2\frac{(\omega-k_\parallel v_\parallel)^2}{\omega_B^2}
\ll 1 \label{eq:32}
\end{equation}
allows us to use so-called drift approximation \cite{TS}.

As it is well known \cite{GS} if turbulence is weak the dynamics
is governed by three wave resonant processes. In $e^-e^+$ plasma
there exists additional limitations for realization of three wave
resonant interactions. The absence of girotropy causes the fact
that the terms proportional to the odd power of the charge do not
contribute to the nonlinear currents. Another specific property of
the magnetosphere plasma is that all the waves are propagating
almost along the magnetic field.

Breakup of $l$ wave into two $t$ waves was studied in \cite{M80},
and all possible three wave processes in \cite{GM83}. In this
researches main attention was payed to possible breakup of $l$
waves because it was believed that these waves should be the most
unstable ones. Existence of the anomalous Doppler-effect resonance
discussed in the previous Section leads to the conclusion that in
fact the most interesting three wave processes are possible
breakups of $t$ waves that should be responsible for the formation
of observed spectra of pulsars.

For the analysis of nonlinear processes it is useful to introduce
quantities and notations used in quantum physics. Usually plasmon
occupation numbers $N_k^\lambda$ and wave amplitudes $a_k^\lambda$
are introduced \cite{ZLF,TS,GS}:
\begin{equation}
\omega_kN_k^\lambda \equiv \omega_k <a_k^\lambda
{a_k^{\lambda}}^\ast>=\frac{1}{\omega_k} \left(
\frac{\partial}{\partial\omega}\omega^2\varepsilon_k^\lambda
\right)_{\omega=\omega_k^\lambda}\frac{|E_k|^2}{4\pi}.
\label{eq:33}
\end{equation}
here
\begin{equation}
\varepsilon_k^\lambda \equiv \frac{{e_i^\lambda}^\ast
e_j^\lambda}{\omega^2}\left( k_ik_jc^2+\omega^2
\varepsilon_{ij}\right); \label{eq:34}
\end{equation}
${\bf e}^\lambda$ is polarization vector of $\lambda (lt,l,t)$
mode; $E_j^\lambda \equiv e_j^\lambda E$; $\varepsilon_{ij}$ is
dielectric tensor; $\ast$ denotes complex conjugation and angular
brackets denote averaging over phases.

Maxwell equations yield
\begin{eqnarray}
\begin{array}{ll}
\partial_t a_k+i\omega_k^\lambda = \sum_{\lambda_1 \lambda_2} \int
d{\bf k}_1d{\bf k}_2d\omega_1d \omega_2 a_k^{\lambda_1}
{a_k^{\lambda_2}}^\ast \\
\times \delta({\bf k}-{\bf k}_1-{\bf k}_2)
\delta(\omega-\omega_1-\omega_2) V_{\lambda |\lambda_1 \lambda_2},
\end{array}\label{eq:35}
\end{eqnarray}
where matrix element of the process $\lambda \rightarrow
\lambda_1+\lambda_2$
\begin{eqnarray}
\begin{array}{ll}
V_{\lambda |\lambda_1 \lambda_2}=4\pi
\frac{\omega_{k_1}^{\lambda_1}
\omega_{k}^{\lambda}}{\omega_{k}^{\lambda}}
(\omega_{k_1}^{\lambda_1}-\omega_{k_2}^{\lambda_2})
(\partial_\omega \omega^2
\varepsilon^\lambda)^{-1/2}_{\omega=\omega_k^\lambda}\\
\times (\partial_\omega \omega^2
\varepsilon^\lambda)^{-1/2}_{\omega=\omega_{k_1}^{\lambda_1}}
(\partial_\omega \omega^2
\varepsilon^\lambda)^{-1/2}_{\omega=\omega_{k_2}^{\lambda_2}}
\sigma_{\lambda |\lambda_1 \lambda_2},
\end{array}\label{eq:36}
\end{eqnarray}
and $\sigma_{\lambda |\lambda_1 \lambda_2}$ is nonlinear
conductivity tensor.

If $k_\perp=0$, the process $t\rightarrow t^\prime + t^{\prime
\prime}$ is forbidden due to the fact that the second order
current $j^{(2)}\sim e^3$. When $k_\perp\neq0$, this process can
take place but the matrix element of interaction is negligibly
small \cite{GM83}: $V_{t|t^\prime t^{\prime \prime}} \sim
\omega_p^3/\omega_B^3\ll 1$.

\begin{figure}[b]
\includegraphics[]{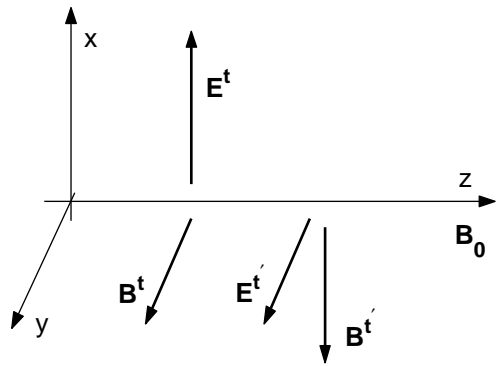}
\caption{\label{fig:fig3} Calculation of nonlinear currents}
\end{figure}

Another possible three wave process $t \rightarrow t^\prime +l$
was discussed in \cite{GM83} without taking into consideration
drift motion of the particles. In this case the frequency of
generated Langmuir waves $\omega_l \approx k_lc$. If all the waves
are propagating along ${\bf B}_0$, matrix element of interaction
is
\begin{equation}
V_{t|l t^\prime} \approx \frac{e}{m_ec} \frac{\omega_p^2}{\omega_B
\omega_l} \sqrt{\frac{\omega_t \omega_{t^\prime}}{\omega_l}}.
\label{eq:310}
\end{equation}
Consideration of the drift motion of particles makes this process
even more intense. For simplicity let us consider the case when
the waves are propagating along ${\bf B}_0$. Let us assume that
electric vectors of $t$ and $t^\prime$ waves are directed along
$x$ and $y$ axes respectively (see Fig. \ref{fig:fig3}) and ${\bf
B}_0$ is parallel to $z$ axis. $E^t_x$ causes the drift motion of
particles along $y$:
\begin{equation}
u_y^d=c\frac{E_x^t}{B_0}, \label{eq:311}
\end{equation}
which on its turn generates nonlinear electric field:
\begin{equation}
{\bf E}=\frac{1}{c}\left[{\bf u}^d, {\bf B}^\prime \right].
\label{eq:312}
\end{equation}
It has to be noted that drift velocity is the same for electrons
and positrons. Consequently the drift motion do not lead to
current generation in the linear approximation. If the resonant
conditions
\begin{equation}
\omega_t=\omega_l+\omega_{t^\prime},~~~{\bf k}_t={\bf k}_l+{\bf
k}_{t^\prime}, \label{eq:313}
\end{equation}
are fulfilled the beating of $t$ and $t^\prime$ waves that
generates longitudinal electric field (\ref{eq:312}) are in
resonance with corresponding $l$ wave.

Analysis of the resonant conditions (\ref{eq:313}) yields that $l$
wave should have phase velocity a bit less then the speed of
light. But this king of $l$ waves in general can be strongly
damped by collisionless Landau damping. In this case instead of
three wave nonlinear interaction one has to consider scattering of
$t$ waves by plasma particles - so called wave-particle-wave
interaction \cite{GS,TS}.

Analysis of standard pulsar parameters yield that in principle
both cases are possible. In this paper we consider the case when
$l$ waves involved in resonance with $t$ waves are not strongly
damped by collisionless damping.

Maxwell equations governing the dynamics in the case when all the
waves are propagating along  ${\bf B}_0$ yield:
\begin{equation}
\left( \frac{\partial^2}{\partial t^2} - c^2
\frac{\partial^2}{\partial z^2} \right){\bf E}_\perp +4\pi
\frac{\partial {\bf j}_\perp^{NL}}{\partial t}=0 ,\label{eq:317}
\end{equation}
\begin{equation}
\frac{\partial E_z}{\partial t} + 4\pi j_z^{NL}=0. \label{eq:318}
\end{equation}
Nonlinear currents ${\bf j}_\perp^{NL}$ ¨ $j_z^{NL}$ can be
readily calculated as follows: substituting (\ref{eq:311}) into
Eq. (\ref{eq:312}) and taking into account Faraday law we get:
\begin{equation}
j_z^{NL}=-\frac{i}{4\pi}\left( \frac{k_z^t}{\omega^t}E_x
E_y^\prime+ \frac{k_z^{t^\prime}}{\omega^{t^\prime}}E_x^\prime E_y
\right). \label{eq:319}
\end{equation}
For ${\bf j}_\perp^{NL}$ we have
\begin{equation}
{\bf j}_\perp^{NL}= {\bf u}_\perp^d \rho_e,\label{eq:320}
\end{equation}
where perturbation of charge density $\rho_e$ can be determined
from Poisson equation.

Substituting in Eqs. (\ref{eq:317})-(\ref{eq:318}) electric field
components as
\begin{equation}
E_{z,\perp}(t)=E_{z,\perp k}(t)e^{i(\omega t-kz)}+c.c.,
\label{eq:321}
\end{equation} and taking into account Eqs.
(\ref{eq:319})-(\ref{eq:320}) as well as dispersions (\ref{eq:21})
and (\ref{eq:242}) we obtain for slowly varying amplitudes:
\begin{equation}
\frac{\partial E_{x k_t}}{\partial t} = \frac{i}{2}
\frac{c\omega_l}{B_0} \left(
\frac{k^{t^\prime}}{\omega^{t^\prime}}+ \frac{k^l}{\omega^l}
\right) E_{z k_l} E_{y k_{t^\prime}}^\ast, \label{eq:322}
\end{equation}
\begin{equation}
\frac{\partial E_{y k_{t^\prime}}}{\partial t} = -\frac{i}{2}
\frac{c\omega_l}{B_0} \left( \frac{k^t}{\omega^t}+
\frac{k^l}{\omega^l} \right) E_{z k_l} E_{x k_t}^\ast,
\label{eq:323}
\end{equation}
\begin{equation}
\frac{\partial E_{z k_l}}{\partial t} = -\frac{i}{2}
\frac{c\omega_l}{\langle \gamma \rangle B_0} \left(
\frac{k^t}{\omega^t}+ \frac{k^{t^\prime}}{\omega^{t^\prime}}
\right) E_{x k_t} E_{y k_{t^\prime}}^\ast. \label{eq:324}
\end{equation}
Assuming that there exist many waves with chaotic phases and
introducing  occupation numbers:
\begin{equation}
N_{x,y k}=\frac{|E_{x,y k}|^2}{4\pi \omega_t},~~~N_{z
k}=\frac{\langle\gamma^2\rangle |E_{zk}|^2}{8\pi \omega_l},
\label{eq:325}
\end{equation}
and using standard technique \cite{TS,GS} we get for the matrix
element of interaction:
\begin{equation}
V_{t|l t^\prime} \approx \frac{c}{B_0} (\omega_t k_{t^\prime}+
\omega_{t^\prime} k_t) \sqrt{\frac{\omega_l}{\omega_t
\omega_{t^\prime}\langle \gamma^2 \rangle}}. \label{eq:326}
\end{equation}
Comparison of Eqs. (\ref{eq:310}) and (\ref{eq:326}) yields that
the drift motion of particles leads to important intensification
of the considered process.

\section{Stationary spectra}

In the case under consideration $\omega_t, \omega_{t^\prime}\gg
\omega_l$. Consequently, the nonlinear interaction is local - only
the waves with close frequencies can effectively interact. As it
was discussed in Sec. 2 the generation of $t$ waves takes place
for $\omega\ll\omega_B$, whereas the cyclotron damping is
effective for $\omega > 2\gamma_p\omega_B$. Nonlinear processes
are responsible for the formation of stationary spectrum between
these intervals of generation and silk. As it is known \cite{ZK78}
the stationary spectrum in the Kolmogorov interval is fully
determined by the matrix element (\ref{eq:326}) and linear
dispersions of waves. There exist two stationary solutions:
\begin{equation}
N_{k1}=A_1 k^{-s-d},~~~N_{k2}=A_2k^{-s-d+1/2}, \label{eq:327}
\end{equation}
where $d$ is the dimension of the problem and $s$ is the index of
homogeneity of the matrix element. The first solution $N_{k1}=A_1
k^{-s-d}$ corresponds to constant flux of energy to the smaller
scales, whereas $N_{k2}=A_2k^{-s-d+1/2}$ corresponds to constant
flux of plasmons to the larger scales \cite{ZLF,ZK78}.

In our case $d=1$ and $S=3/2$. Therefore $N_{k1}=A_1 k^{-5/2}$ and
$N_{k2}=A_2 k^{-2}$. Taking into account the linear character of
$t$ wave dispersion for the energy spectrum ${\cal E}_{\omega}$
that is proportional to the observed intensity $I_\omega$ we have
\begin{equation}
I_{\omega 1} \sim {\cal E}_{\omega 1} \sim
\omega^{-3/2},~~~I_{\omega 2} \sim {\cal E}_{\omega 2} \sim
\omega^{-1}. \label{eq:328}
\end{equation}

which of this spectra is realized in practice depends on the
linear mechanisms of generation and silk of the waves. In the case
under consideration the waves with relatively low frequencies are
excited at the anomalous doppler-effect resonance and waves with
relatively high frequencies are damped at cyclotron resonance. In
this case the spectrum ${\cal E}_{\omega 1} \sim \omega^{-3/2}$ is
only possible.

As it was mentioned above most of the pulsars has the spectral
indexes $-1.5>\alpha>-2$. Consequently, obtained theoretical
result $\alpha=-1.5$ seems to be in good accordance with the
observations.

\section{Summary}

Nonlinear three wave processes that can be responsible for the
formation of spectra of radio pulsars is considered. Existence of
strong magnetic field makes it necessary to take into account
electric drift motion of particles. In the framework of the weak
turbulence theory possible stationary spectra that can occur in
Kolmogorov interval are calculated. Analysis of linear mechanisms
of wave generation and silk allows to conclude that only possible
spectral index is $\alpha =-1.5$. Obtained result is in
satisfactory accordance with the observational data.

\end{document}